\def\la{\mathrel{\hbox{\rlap{\hbox{\lower4pt\hbox{$\sim$}}}\hbox{$<$}}}}
\def\ga{\mathrel{\hbox{\rlap{\hbox{\lower4pt\hbox{$\sim$}}}\hbox{$>$}}}}

\def\Teff{\ifmmode{T_{\rm eff}}\else{\hbox{$T_{\rm eff}$} }\fi}
\def\Rzero{\ifmmode{R_0}\else{\hbox{$R_0$} }\fi}
\def\kms{km s$^{-1}$}

\def\56ni{$^{56}$Ni}
\def\56co{S^{56}$Co}

\def\vph{$v_{phot}}

\documentstyle[12pt,aasms4]{article}

\received{}
\accepted{}
\journalid{}{}
\articleid{}{}
\begin{document}
\title {Direct Analysis of Spectra of Type~Ic Supernovae}

\author {David Branch}

\affil {Department of Physics and Astronomy, University of
Oklahoma, Norman, Oklahoma 73019, USA}

\begin{abstract}

Synthetic spectra generated with the parameterized supernova
synthetic--spectrum code SYNOW are compared with observed
photospheric--phase optical spectra of the normal Type~Ic SN~1994I and
the peculiar Type~Ic SNe 1997ef and 1998bw.  The observed spectra can
be matched fairly well with synthetic spectra that are based on
spherical symmetry and that include lines of just a few ions that are
expected to appear on the basis of LTE calculations.  Spectroscopic
estimates of the mass and kinetic energy of the line--forming layers
of the ejected matter give conventional values for SN~1994I but high
kinetic energy ($\sim 30
\times 10^{51}$ erg) for SN~1997ef and even higher ($\sim 60
\times 10^{51}$ erg) for SN~1998bw.  It is likely that even if SNe~1997ef
and 1998bw were non--spherical, they also were hyper--energetic.

\end{abstract}

\section{Introduction}

The photospheric--phase spectrum of a Type~Ic supernova lacks the
strong hydrogen lines of a Type~II, the strong optical He~I lines of a
Type~Ib, and the deep red Si~II absorption of a Type~Ia.  A Type~Ic
(SN~Ic) is thought to be the result of the core collapse of a massive
star that either has lost its helium layer or ejects helium that
remains insufficiently excited to produce conspicuous optical He~I
lines.  For a review of observations of supernova spectra, see
Filippenko (1997).  

Since April, 1998, interest in SNe~Ic has been very high because of
the extraordinary SN~1998bw, which appears to have been associated
with the gamma ray burst GRB 980425 (Galama et~al. 1998; Kulkarni
et~al. 1998).  Recently we (Millard et~al. 1999) have used the
parameterized supernova synthetic--spectrum code SYNOW to make a
``direct analysis'' of spectra of SN~1994I, a well observed, normal
SN~Ic.  In this contribution, after summarizing our work on SN~1994I,
I report some preliminary results of a similar analysis of the
peculiar Type~Ic SN~1997ef (Deaton et~al. 1998; J.~Millard et~al., in
preparation) and then consider the related but even more peculiar
SN~1998bw.  The emphasis here is on establishing line identifications
and making spectroscopic estimates of the mass and kinetic energy of
the line--forming layers of the ejected matter.

\section{SYNOW}

In its simplest form the SYNOW code assumes spherical symmetry and
that line formation takes place by resonant scattering outside a sharp
photosphere that radiates a blackbody continuum.  To a good
approximation the simple explosion velocity law, $v=r/t$, holds.
Consequently the velocity gradient is isotropic and homogeneous
(unlike the case of a stellar wind, where even in the
constant--velocity case the velocity gradient is neither isotropic nor
homogeneous).  From the point of view of an observer, the
(nonrelativistic) surfaces of constant radial velocity are planes
perpendicular to the line of sight, and an unblended line formed by
resonant scattering has an emission component that peaks at the line
rest wavelength in the supernova frame and an absorption component
whose minimum is blueshifted by an amount that corresponds to the
velocity at the photosphere (unless the line is extremely weak or
rather strong).

SYNOW treats line formation in the Sobolev approximation, which is a
good one for this purpose.  The profile of an unblended line is
determined by the adopted radial dependences of the line optical depth
and the line source function.  The line optical depth determines the
strength of the line and is given by

$$ \tau_l = {{\pi e^2} \over {m_e c}}\ f\ \lambda\ t\ n_l(v) = 0.026\
f\ \lambda_\mu\ t_d\ n_l(v), $$

\noindent where $f$ is the oscillator strength, $\lambda_\mu$ is the 
line wavelength in microns, $t_d$ is the time since explosion in days,
and $n_l(v)$ is the population of the lower level of the transition in
cm$^{-3}$.  The correction for stimulated emission, although not
written out here, is taken into account.

The line source function determines the extent to which the line is in
emission or absorption.  In the resonant--scattering approximation
line photons are conserved, except for occultation effects, and the
source function of an isolated line is just the product of the
intensity of the photospheric continuum and the geometrical dilution
factor.  The source function of a line that interacts with one or more
lines of shorter wavelength is altered by photons that are scattered
by those lines.  The essential role of SYNOW is to treat the multiple
scattering, which in observers' language is line blending.

Various fitting parameters are available for a SYNOW
synthetic--spectrum calculation.  $T_{bb}$ is the temperature of the
blackbody continuum radiated by the photosphere.  For each ion whose
lines are introduced, the optical depth at the photosphere of a
reference line is a parameter, and the optical depths of the other
lines of the ion are calculated for Boltzmann excitation at excitation
temperature $T_{exc}$ (which ordinarily is taken to be the same as
$T_{bb}$).  For the spectra shown here, the radial dependence of the
line optical depths is a power law, $\tau \propto v^{-n}$.  At each
epoch, the velocity at the photosphere, $v_{phot}$, is a parameter,
and maximum and minimum velocities also can be imposed on an ion; when
the minimum velocity exceeds $v_{phot}$ the ion is said to be detached
from the photosphere.  The most interesting parameters are the
velocity parameters and the ``density'' power--law index, $n$.

When deciding which ions to introduce, we are guided by experience and
by the LTE calculations of line optical depths by Hatano
et~al. (1999), who considered six compositions that might be
encountered in supernovae.  The composition of interest here is the
one in which hydrogen and helium have been burned to a mixture of
carbon and oxygen, with the heavier elements present in their solar
mass fractions.  In this case, ions that are predicted to have lines
of significant optical depth include Ca~II, Fe~II, O~I, Si~II, C~II,
Mg~II, O~II, and Ti~II.

\section{The Normal Type~Ic SN~1994I}

In Millard et~al. (1999) we compare SYNOW synthetic spectra with
observed spectra of SN~1994I obtained by Filippenko et~al. (1995) from
5 to 35 days after the assumed explosion date of March~30, 1994.  A
density power--law index of $n=8$ is used for all epochs.  Figure~1
compares a spectrum of SN~1994I obtained 16 days after explosion with
a synthetic spectrum that has $v_{phot}=10,000$~\kms\ and
$T_{bb}=8000$~K.  Most of the observed features are well matched.
Ions that certainly must be introduced to account for observed
features are Ca~II, O~I, Na~I, Fe~II, and Ti~II.  (The Na~I D--line
feature is not predicted to be significant by Hatano et~al. (1999),
but as usual in supernova spectra the Na~I feature is observed to be
stronger than predicted.  We are confident of the identification.)  In
this particular synthetic spectrum, lines of C~II also are used, but
detached at 16,000
\kms\ so that $\lambda6580$ can account for most of the observed
absorption near 6200~\AA.  Often it is difficult to decide between
detached C~II $\lambda6580$ and undetached Si~II $\lambda6355$, and
the C~II identification is not considered to be definite.  In order to
account for the observed absorption around 7000~\AA\ we would have to
introduce lines of O~II (see below).  The excessive height of the
synthetic peaks in the blue part of the spectrum is not of great
concern; the number of lines of singly ionized iron--peak elements
rises rapidly toward short wavelengths, so SYNOW spectra often are
underblanketed in the blue due to missing lines of iron--peak ions
that are not introduced.

In Millard et~al. (1999) we consider the identification of the
observed absorption near 10,250~\AA, which has been identified as He~I
$\lambda10830$ and taken as strong evidence that SN~1994I ejected
helium (Filippenko et~al. 1995).  We find that it is difficult to
account for even just the core of the observed 10,250~\AA\ absorption
with He~I $\lambda$10830 without compromising the fit in the optical
(see also Baron et~al. 1999).  We suggest that the observed feature
may be a blend of He~I $\lambda$10830 and C~I $\lambda$10695, or
perhaps a blend of Si~I lines.  This is an important issue but it
will not be discussed further here, since we have no evidence for or
against the presence of the feature in SNe~1997ef and 1998bw.

Another comparison of observed and synthetic spectra, but for just
five days after the assumed explosion date, is shown in Figure~2.  (We
often consider photospheric--phase spectra in reverse chronological
order, because at the earliest times line formation takes place in the
highest--velocity layers and the blending is most severe.)  The
synthetic spectrum has $v_{phot}=17,500$
\kms\ and $T_{bb}=17,000$~K.  
Ions that certainly are needed are Ca~II, O~I, and Fe~II.  In this
synthetic spectrum, lines of C~II, Na~I, Mg~II, Si~II, and O~II (with
only [O~II] $\lambda$7320,7330 having a significant effect on the
spectrum) also are introduced; they are considered probable but not
definite.

The adopted values of $v_{phot}$ can be used to estimate the mass and
kinetic energy in the line--forming layers.  For an $r^{-n}$ density
distribution, the mass (in $M_\odot$) and the kinetic energy (in foe,
where 1 foe $\equiv 10^{51}$ erg) above the layer at which the
electron--scattering optical depth is $\tau_{es}$ can be expressed as
$$ M = 1.2 \times 10^{-4}\ v_4^2\ t_d^2\ \mu_e\ \tau_{es}\ f_M(n, v_{max}),
$$
$$ E = 1.2 \times 10^{-4}\ v_4^4\ t_d^2\ \mu_e\ \tau_{es}\ f_E(n,
v_{max}), $$
\noindent where $v_4$ is $v_{phot}$ in units of $10^4$ \kms, $t_d$ is the time
since explosion in days, $\mu_e$ ($\equiv Y_e^{-1}$) is the mean
molecular weight per free electron, and the integration is carried out
to velocity $v_{max}$.  The functions $f_M$ and $f_E$ always exceed
unity.

For SN~1994I, we use $\mu_e=14$ (a mixture of singly ionized carbon
and oxygen), $\tau_{es}=2/3$ at the bottom of the line--forming layer,
and integrate the steep density power--law to infinity ($f_M(8,
\infty)=1.4$, $f_E(8, \infty)=2.3$).  Figure~3 shows $v_4$, $M$,
and $E$ plotted against time.  ($E$ should increase monotonically with
time. Its non--monotonic behavior just reflects the imprecision of our
determinations of $v_{phot}$; recall that $E \propto {v_{phot}^4}$.)
At 35 days after explosion, the mass moving faster than 7000~\kms\ is
estimated to be about 1.4~$M_\odot$ and it carries a kinetic energy of
about 1.2~foe.  These numbers are reasonable, and similar to those
that have been estimated for SN~1994I on the basis of light--curve
studies (Nomoto et~al. 1994; Iwamoto et~al. 1994; Young, Baron, \&
Branch 1995; Woosley, Langer, \& Weaver 1995).

Such spectroscopic estimates of mass and kinetic energy also come out
to be reasonable for Type~Ia supernovae (Branch 1980) and for SN~1987A
(Jeffery \& Branch 1990).

\section {From SN~1994I to SNe~1997ef and 1998bw}

Figure~4 compares spectra of SN~1994I at 16 days after explosion,
SN~1997ef at 20 days after its assumed explosion date of November 15,
1997, and SN~1998bw at 16 days after its explosion date of April 25,
1998.  It is clear that SNe~1997ef and 1998bw are spectroscopically
related to each other and also, but less closely, to SN~1994I.
Therefore it seems appropriate to refer to SNe~1997ef and 1998bw as
``Type~Ic peculiar''.  The observed absorption features are much
broader and bluer in SN~1997ef than in SN~1994I, and even moreso in
SN~1998bw.  This means that SNe~1997ef and 1998bw ejected more mass at
high velocity.  Figure~5 shows the effects, on the SYNOW synthetic
spectrum of Figure~1 (for SN~1994I at 16 days), of raising $v_{phot}$
from 10,000 to 30,000~\kms. Figure~6 shows the effects of dropping the
density power--law index from $n=8$ to $n=2$.  Raising $v_{phot}$ and
dropping $n$ both cause the absorption features to become broader and
bluer, and both appear to be necessary to obtain satisfactory SYNOW
fits to spectra of SNe~1997ef and 1998bw.  A value of $n=2$ is used
for all of the synthetic spectra shown below.

\subsection{Fitting spectra of SN~1997ef}

Figure~7 compares a spectrum of SN~1997ef obtained 34 days after
explosion with a synthetic spectrum that has \vph=7000$ \kms, $T_{bb}
= 7000$~K, and uses only lines of Ca~II, O~I, Si~II, and Fe~II.  This
fit (and those to follow) could be improved by tuning the parameters,
but as it stands it is good enough to indicate that we are on the
right track.

Figure~8 compares a spectrum of SN~1997ef obtained 20 days after
explosion with a synthetic spectrum that has \vph=12,000$ \kms,
$T_{bb} = 11,000$~K, and uses lines of Ca~II, O~I, Si~II, Fe~II, and
Mg~II.  In the red part of the spectrum, the good fit indicates that
at this epoch just a few lines of Ca~II, O~I, and Si~II are
responsible for the features.  In the blue, Fe~II blends dominate.
The synthetic spectrum is severely underblanketed in the blue due to
missing lines of other singly ionized iron--peak elements.

Figure~9 compares a spectrum of SN~1997ef obtained 10 days after
explosion with a synthetic spectrum that has \vph=22,000$ \kms,
$T_{bb} = 11,000$~K, and uses the same ions as in Figure~8.  This is a
good example of why it can be instructive to work backward in time;
this interpretation of the spectrum might seem arbitrary if the
later--epoch spectra had not already been discussed.

Figure~10 is exactly like Figure~9 except that the Fe~II lines have
been turned off.  Comparison of Figures~9 and 10 shows how strongly
the Fe~II lines affect the blue, while having practically no
effect in the red.  The same is true for the 20 and 34 day spectra
discussed above.

\subsection{Fitting spectra of SN~1998bw}

Figure~11 compares a spectrum of SN~1998bw obtained 28 days after
explosion with a synthetic spectrum that has
\vph=7000$~\kms, $T_{bb}=6000$~K, and uses lines of Ca~II, O~I,
Si~II, Na~I, Ca~II, and Fe~II.  The situation is much like that of
SN~1997ef at 34 days.

Figure~12 compares a spectrum of SN~1998bw obtained 16 days after
explosion with a synthetic spectrum that has \vph=17,000$~\kms,
$T_{bb}=8000$~K, and uses only lines of Ca~II, O~I, Si~II, and Fe~II.
The situation is like that of SN~1997ef at 20 days, but here the
blending is more severe due to the higher $v_{phot}$ of SN~1998bw.

Figure~13 compares a spectrum of SN~1998bw obtained 8 days after
explosion with a synthetic spectrum that has \vph=30,000$~\kms,
$T_{bb}=8000$~K, and uses only lines of O~I, Si~II, Ca~II, and Fe~II.
Again the situation is like that of SN~1997ef at 10 days but with more
blending due to higher $v_{phot}$.

Figure~14 is like Figure~13 except that the Fe~II lines have been
turned off.  

\subsection {Masses and kinetic energies of SNe~1997ef and 1998bw}

Figure~15 compares the adopted values of $v_{phot}$ versus
time. Around 30 days after explosion the $v_{phot}$ values converge to
about 7000~\kms, but at earlier times the values for SN~1997ef are
higher than those for SN~1994I, and the values for SN~1998bw are
higher still.

To estimate the masses and kinetic energies of SNe~1997ef and 1998bw,
the integration cannot be extended to infinity because $n=2$ has been
used for the synthetic spectra.  Instead the integration is carried
out to $v_{max} = 2 v_{phot}$ (with $f_M(2,2)=2, f_E(2,2)=4.7$) for
the earliest epoch considered, i.e., to 44,000~\kms\ for SN~1997ef and
to 60,000~\kms\ for SN~1998bw.  The results are shown in Figures~16
and 17.  For SNe~1997ef and 1998bw, the masses above 7000~\kms\ are
estimated to be around 6 $M_\odot$.  For SN~1997ef the kinetic energy
above 7000~\kms\ comes out to be around 30~foe while that of SN~1998bw
is around 60~foe.  These kinetic energies are more likely to be too
low than too high because most of the estimated kinetic energy comes
from the earliest epochs considered, and the integrations are carried
out to only $2v_{phot}$ while the synthetic spectra actually go to
higher velocities.  Of course, there also is more mass at velocities
lower than 7000~\kms, but not much more kinetic energy.

In a preprint, Iwamoto et~al. (1998) compare observed spectra of
SN~1997ef with synthetic spectra calculated for a hydrodynamical model
that has an ejected mass of about 4.6~$M_\odot$ and a kinetic energy
of 1~foe.  The prominent lines in their synthetic spectra are much the
same as the ones that have been identified here but as they discuss,
the lines in their synthetic spectra are much too narrow and not
sufficiently shifted to the blue.  Synthetic spectra calculated for
models having more mass and kinetic energy give much better fits to
the SN~1997ef spectra (P.~Mazzali and K.~Nomoto, personal
communication).  Iwamoto et~al. (1999) compare observed spectra of
SN~1998bw with synthetic spectra calculated for a hydrodynamical model
that has an ejected mass of about 11~$M_\odot$ and a kinetic energy of
30~foe.  Their synthetic spectra match the SN~1998bw spectra fairly
well, and it appears that more mass at high velocity would lead to
even better fits.

\section{Conclusion}

The spectroscopic mass and kinetic--energy estimates presented here
for SNe~1997ef and 1998bw are preliminary and approximate.
Nevertherless, it seems clear that at least in the spherical
approximation the kinetic energy of both events was much higher than
the canonical one foe, as was reported by Deaton et~al. (1998) for
SN~1997ef and as in the models of Iwamoto et~al. (1999) and Woosley,
Eastman, \& Schmidt (1999) for SN~1998bw.

Polarization spectra are much more sensitive than flux spectra to
asymmetry.  Core--collapse supernovae generally show detectable
polarization, which indicates that they are significantly asymmetric
(Wang et~al. 1996).  H\"oflich, Wheeler, \& Wang (1999) calculate
light curves of moderately asymmetric explosions and suggest that
SN~1998bw was distinguished principally by having been viewed close to
the symmetry axis, rather than by having a very high kinetic energy.
It is true that to the extent that the ejecta of SNe~1997ef and 1998bw
are ``beamed'', the kinetic energy estimates presented here might be
too high.  However, because the spectra of SNe~1997ef and 1998bw can
be matched fairly well in a straightforward way with the spherical
symmetry assumption, and the corresponding kinetic--energy estimates
are so very high, and the Lorentz factor of the ejecta is not high
enough to {\sl be} a factor in the energy estimates, it is likely that
even if SNe~1997ef and 1998bw were non--spherical, they also were
hyper--energetic.

\bigskip

I am grateful to Peter Garnavich and Yulei Qiu for providing spectra
of SN~1997ef, to Ferdinando Patat for providing spectra of SN~1998bw,
and to Eddie Baron, Kazuhito Hatano, David Jeffery, and Jennifer
Millard for discussions and assistance.

\clearpage

\begin {references}

\reference{} Baron, E., Branch, D., Hauschildt, P. H.,
Filippenko,~A.~V., \& Kirshner,~R.~P. 1999, ApJ, submitted

\reference{} Branch, D. 1980, in Proceedings of the Texas Workshop on
Type I Supernovae, ed. J.~C.~Wheeler (Austin: University of Texas),
p. 66

\reference{} Deaton, J., Branch, D., Baron, E., Fisher, A.,
Kirshner,~R.~P., \& Garnavich,~P. 1998, BAAS, 30, 824

\reference{} Filippenko, A. V. 1997, ARAA, 35, 309

\reference{} Filippenko, A. V. et al. 1995, ApJ, 450, L11

\reference{} Galama, T. J. et al. 1998, Nature, 395, 670

\reference{} Hatano, K., Branch, D., Fisher, A., Millard,~J., \&
Baron,~E. 1999, ApJS, 121, 233

\reference{} H\"oflich, P., Wheeler, J. C., \& Wang, L. 1999, preprint

\reference{} Iwamoto, K. et al. 1994, ApJ, 437, L115

\reference{} Iwamoto, K. et al. 1998, astro--ph/9807060

\reference{} Iwamoto, K. et al. 1999, Nature, 395, 672

\reference{} Jeffery, D. J. \& Branch, D. 1990, in Supernovae,
ed. J.~C.~Wheeler, T.~Piran, \& S.~Weinberg (Singapore, World
Scientific), p. 149

\reference{} Kulkarni, S. R. et al. 1998, Nature, 395, 663

\reference{} Millard, J. et al. 1999, ApJ, in press

\reference{} Nomoto, K. et al. 1994, Nature, 371, 227

\reference{} Wang, L., Wheeler, J. C., Li, Z., \& Clocchiatti,
A. 1996, ApJ, 467, 435

\reference{} Woosley, S. E., Eastman, R. G., \& Schmidt,~.B.~P. 1999,
ApJ, in press

\reference{} Woosley, S. E., Langer, N., \& Weaver, T. A. 1995, ApJ,
448, 315

\reference{} Young, T. R., Baron, E., \& Branch, D. 1995, ApJ, 449, L51

\end{references}

\clearpage

\begin{figure} 
\figcaption{A spectrum of SN~1994I obtained 16 days after explosion is 
compared with a synthetic spectrum (dashed line).  The flux is per
unit frequency.}
\end{figure}

\begin{figure} 
\figcaption{A spectrum of SN~1994I obtained 5 days after explosion is 
compared 
with a synthetic
spectrum.  The flux is per unit frequency.}
\end{figure}

\begin{figure} 
\figcaption{Velocity at the
photosphere (in units of 10,000 \kms), and mass (in $M_\odot$) and
kinetic energy (in foe) above the photosphere, are plotted against
time for SN~1994I.}
\end{figure}

\begin{figure} 
\figcaption{A spectrum of SN~1994I (Filippenko et~al. 1995) is compared
with spectra of SN~1997ef (P.~Garnavich et~al., in preparation) and
SN~1998bw (F.~Patat et~al., in preparation).  In this and subsequent
figures the flux is per unit wavelength.}
\end{figure}

\begin{figure} 
\figcaption{The dashed line is the synthetic spectrum of Figure~1.
The solid line shows the effects of raising $v_{phot}$ from 10,000 to
30,000~\kms.}
\end{figure}

\begin{figure} 
\figcaption{The dashed line is the synthetic spectrum of Figure~1. The
solid line shows the effects of dropping $n$ from 8 to 2.}
\end{figure}

\begin{figure} 
\figcaption{A spectrum of SN~1997ef (Y.~Qiu et~al., in
preparation) obtained 34 days after explosion is compared with a
synthetic spectrum.}
\end{figure}

\begin{figure} 
\figcaption{A spectrum of SN~1997ef (P.~Garnavich et~al., in
preparation) obtained 20 days after explosion is compared with a
synthetic spectrum.}
\end{figure}

\begin{figure} 
\figcaption{A spectrum of SN~1997ef (P.~Garnavich et~al., in
preparation) obtained 10 days after explosion is compared with a
synthetic spectrum.           }
\end{figure}

\begin{figure} 
\figcaption{The synthetic spectrum is like that of Figure~9 except
that the Fe~II lines have been turned off.}
\end{figure}

\begin{figure} 
\figcaption{A spectrum of SN~1998bw (F.~Patat et~al., in preparation)
obtained 28 days after explosion is compared with a synthetic spectrum.}
\end{figure}

\begin{figure} 
\figcaption{A spectrum of SN~1998bw (F.~Patat et~al., in preparation)
obtained 16 days after explosion is compared with a synthetic spectrum.}
\end{figure}

\begin{figure} 
\figcaption{A spectrum of SN~1998bw (F.~Patat et~al., in preparation)
obtained 8 days after explosion is compared with a synthetic spectrum.}
\end{figure}

\begin{figure} 
\figcaption{The synthetic spectrum is like that of Figure~13 except
that the Fe~II lines have been turned off.}
\end{figure}

\begin{figure} 
\figcaption{Velocity at the photosphere (in units of 10,000~\kms) is 
plotted against time.}
\end{figure}

\begin{figure} 
\figcaption{Spectroscopic estimates of mass (in $M_\odot$) above the
photosphere are plotted against time.}
\end{figure}

\begin{figure} 
\figcaption{Spectroscopic estimates of kinetic energy (in foe) above the
photosphere are plotted against time.}
\end{figure}

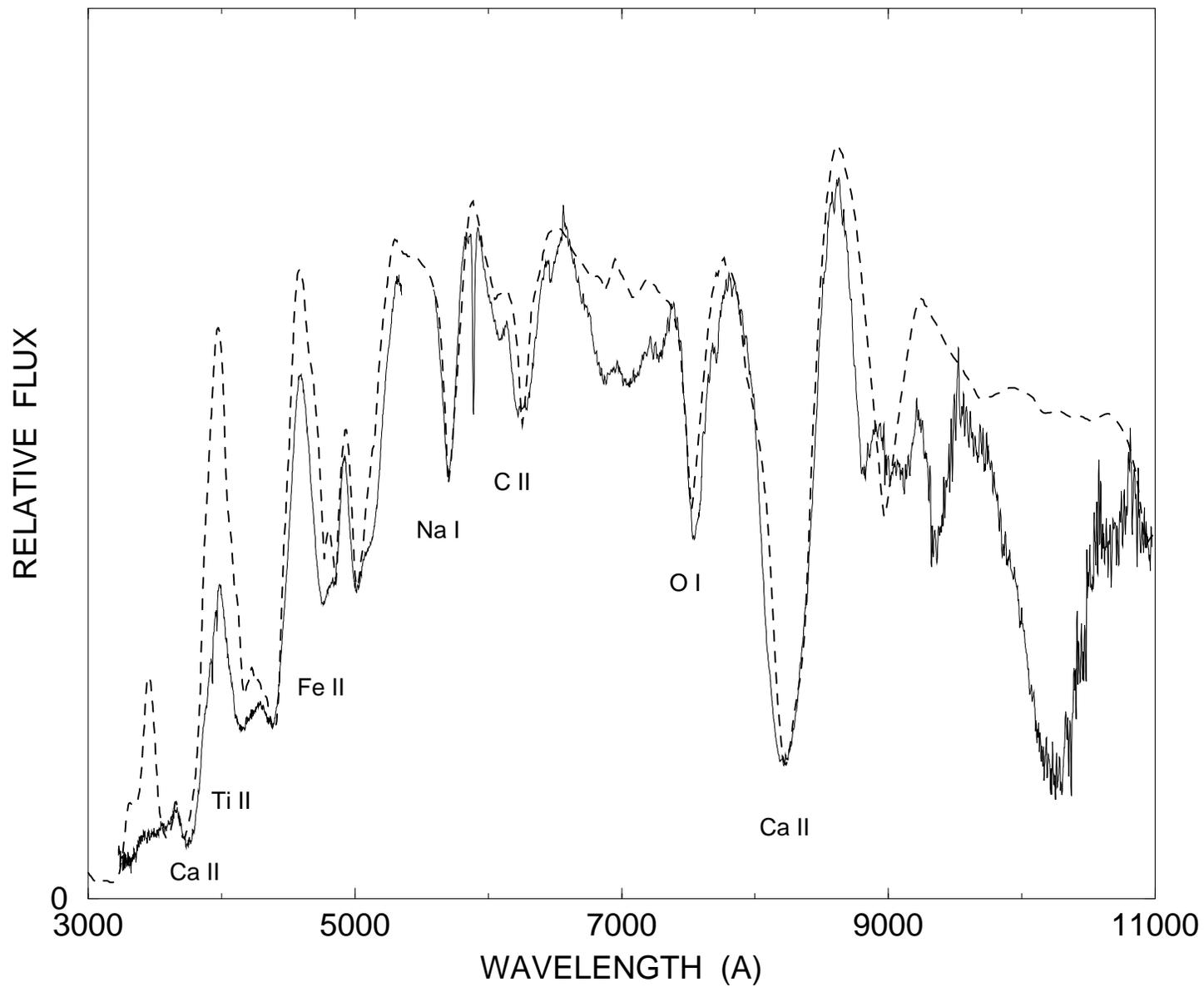
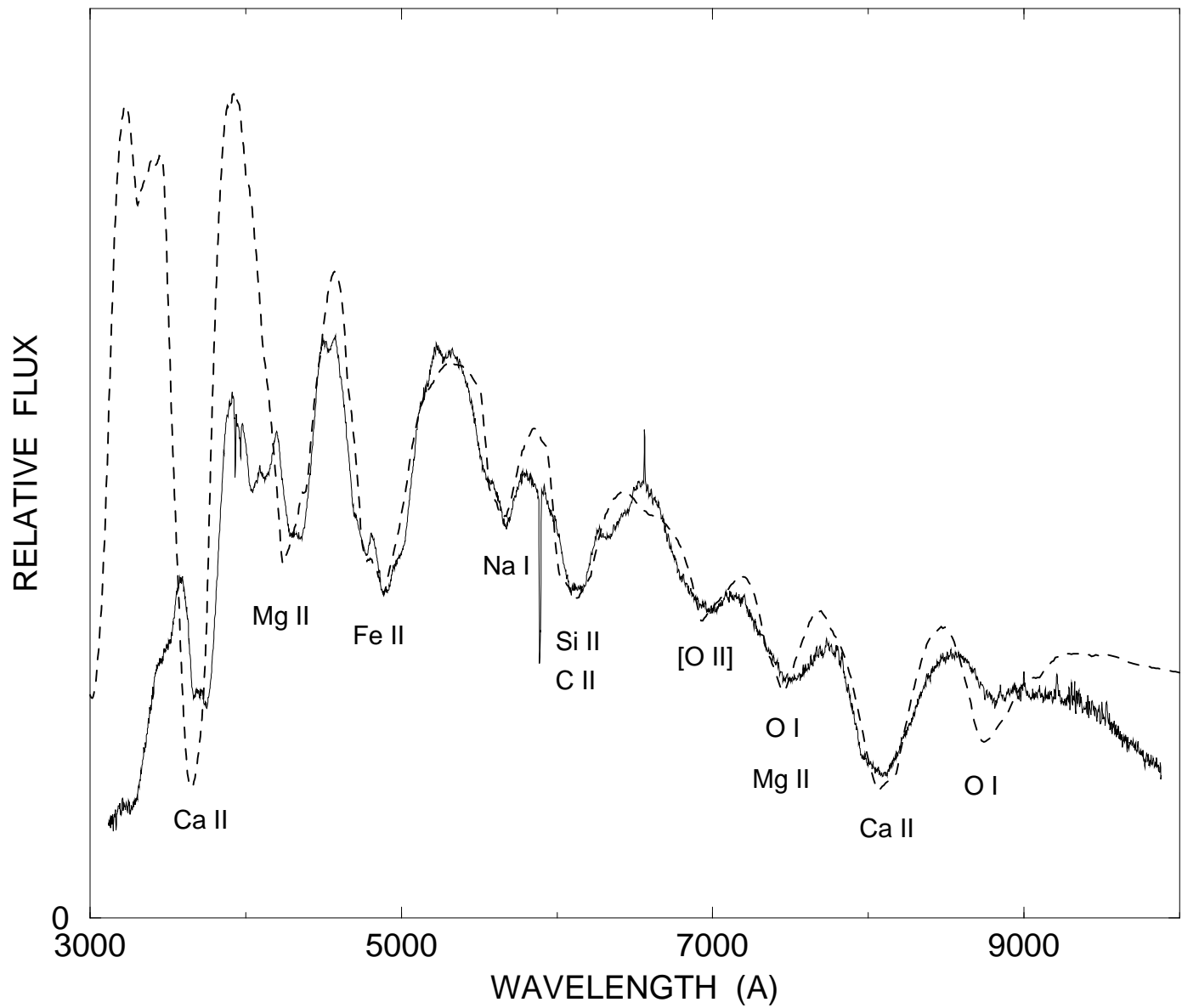
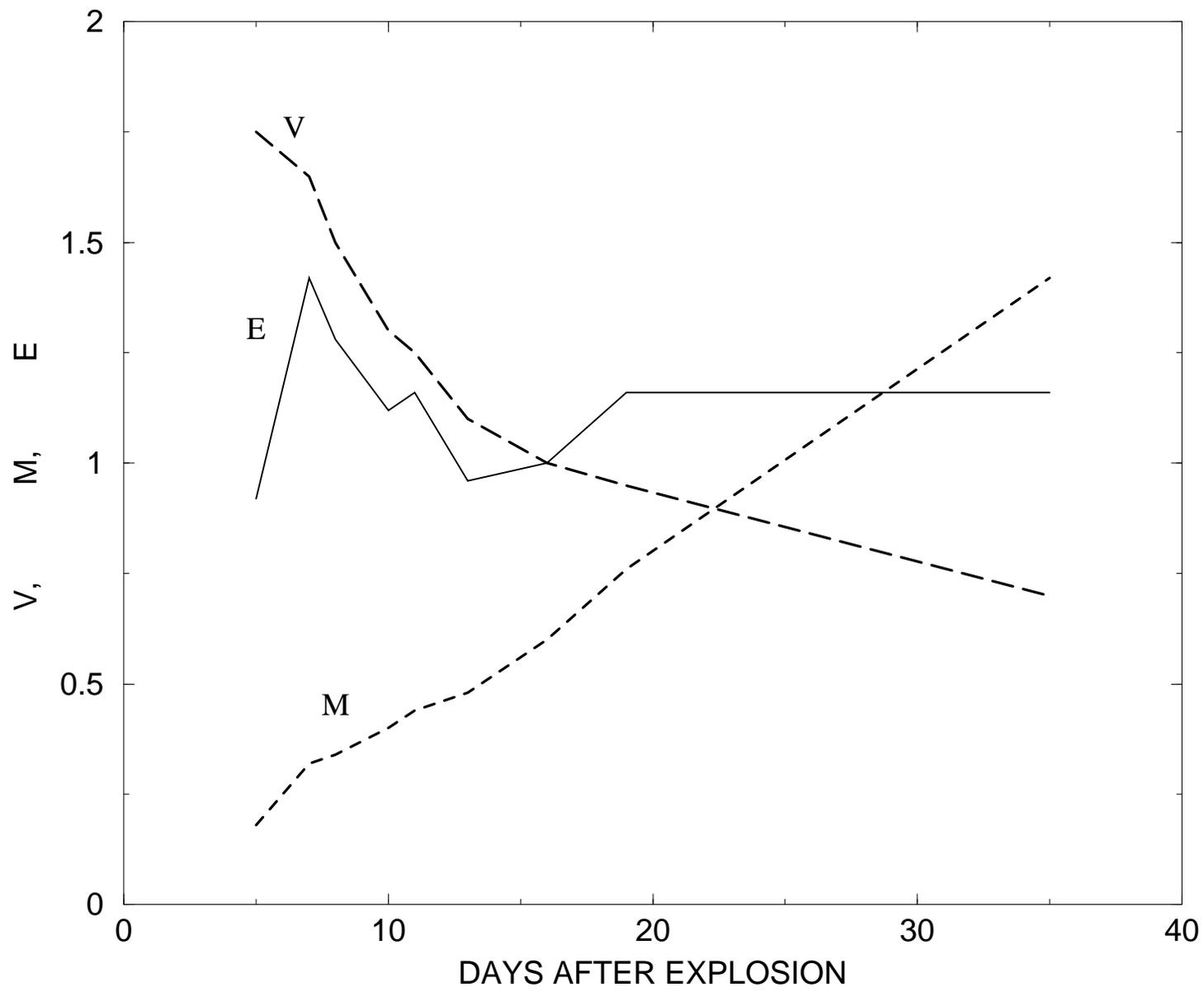
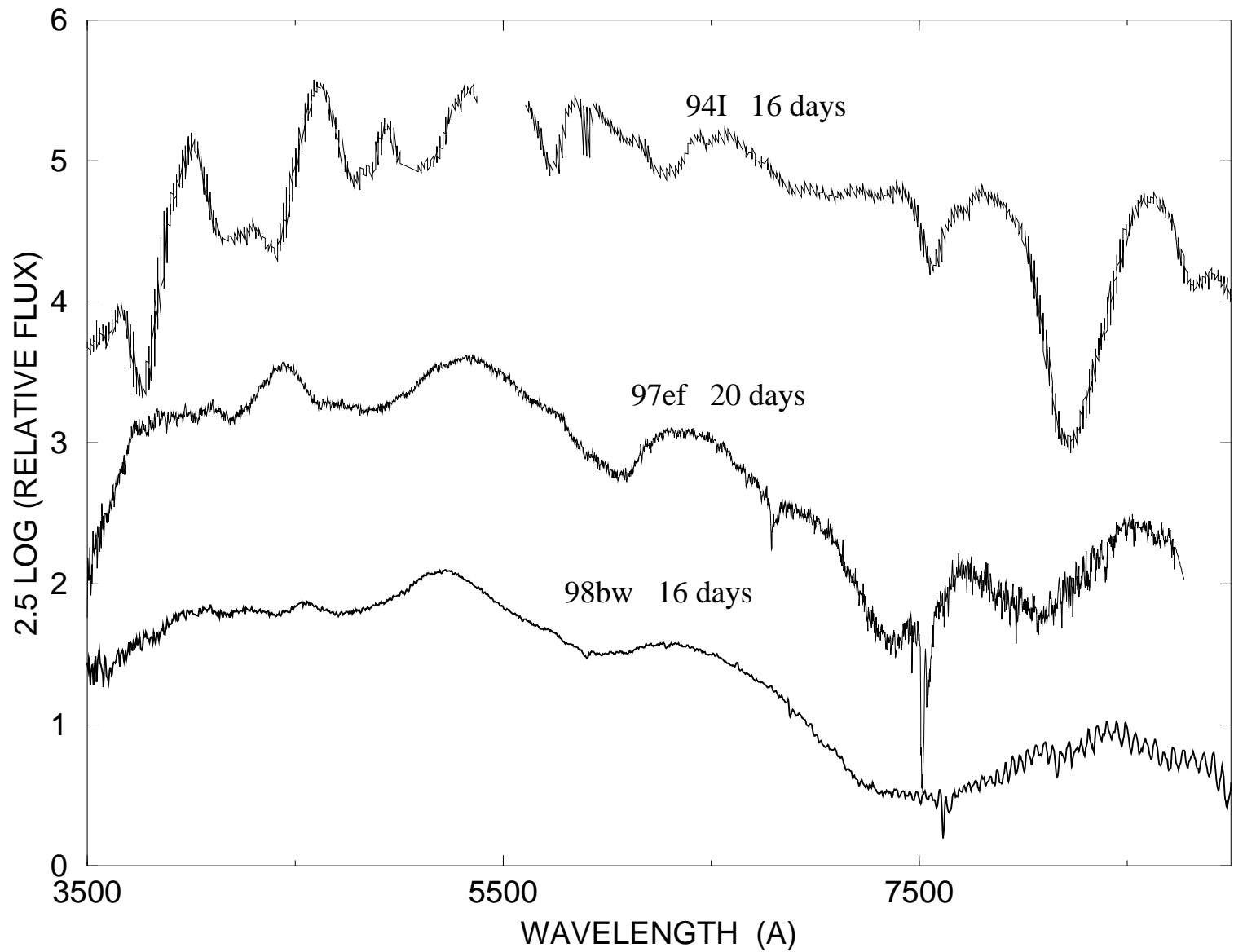
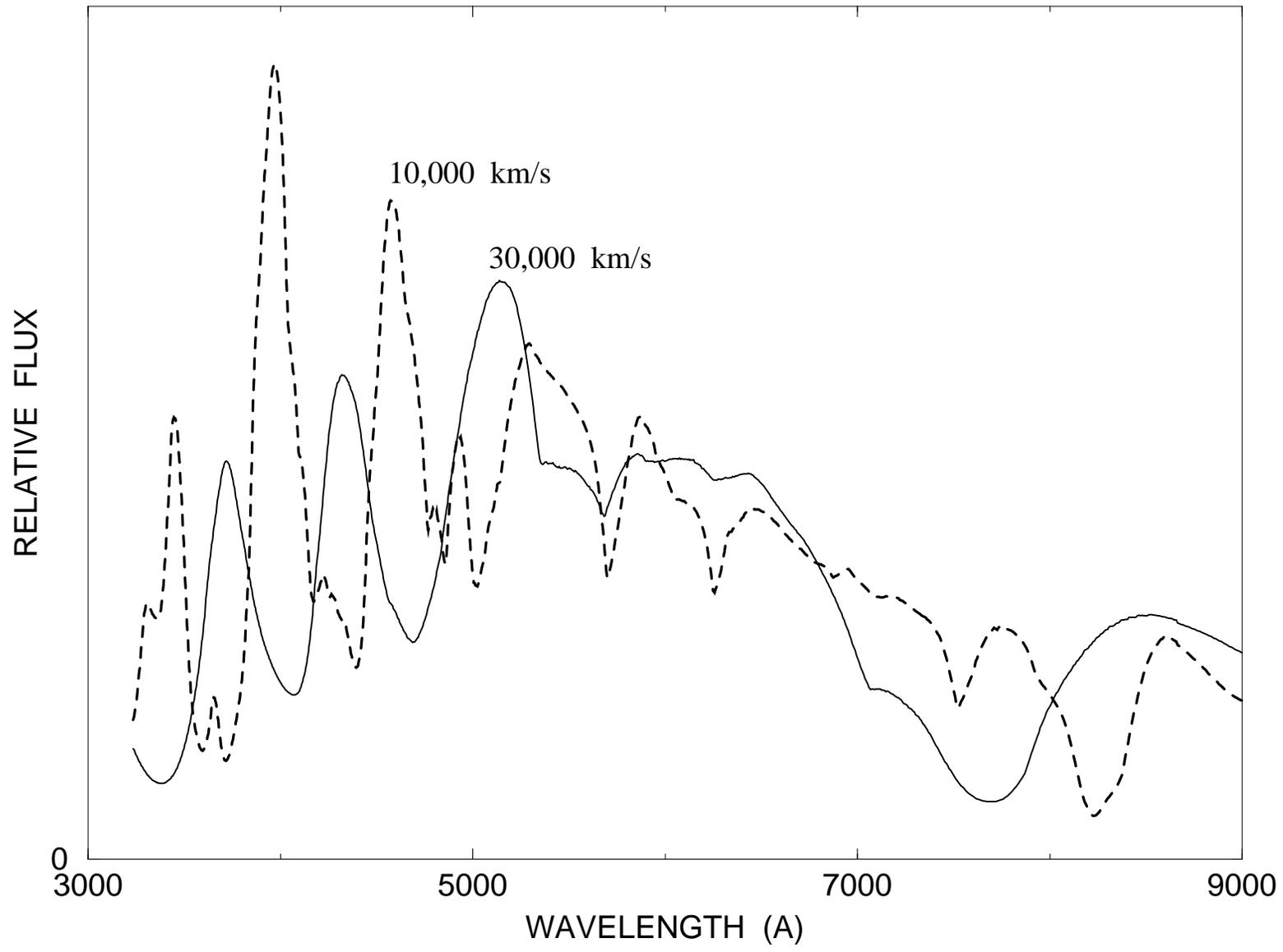
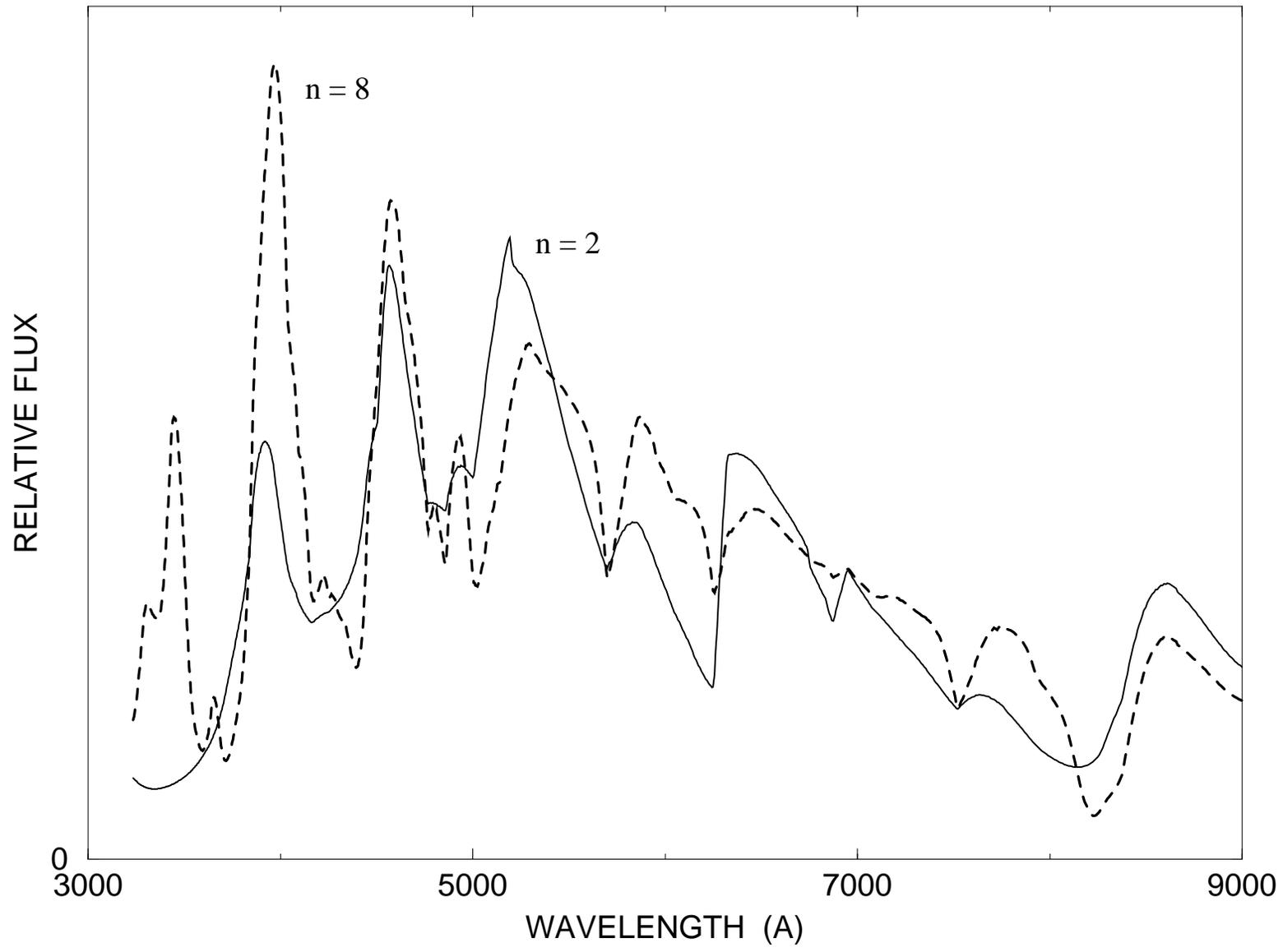
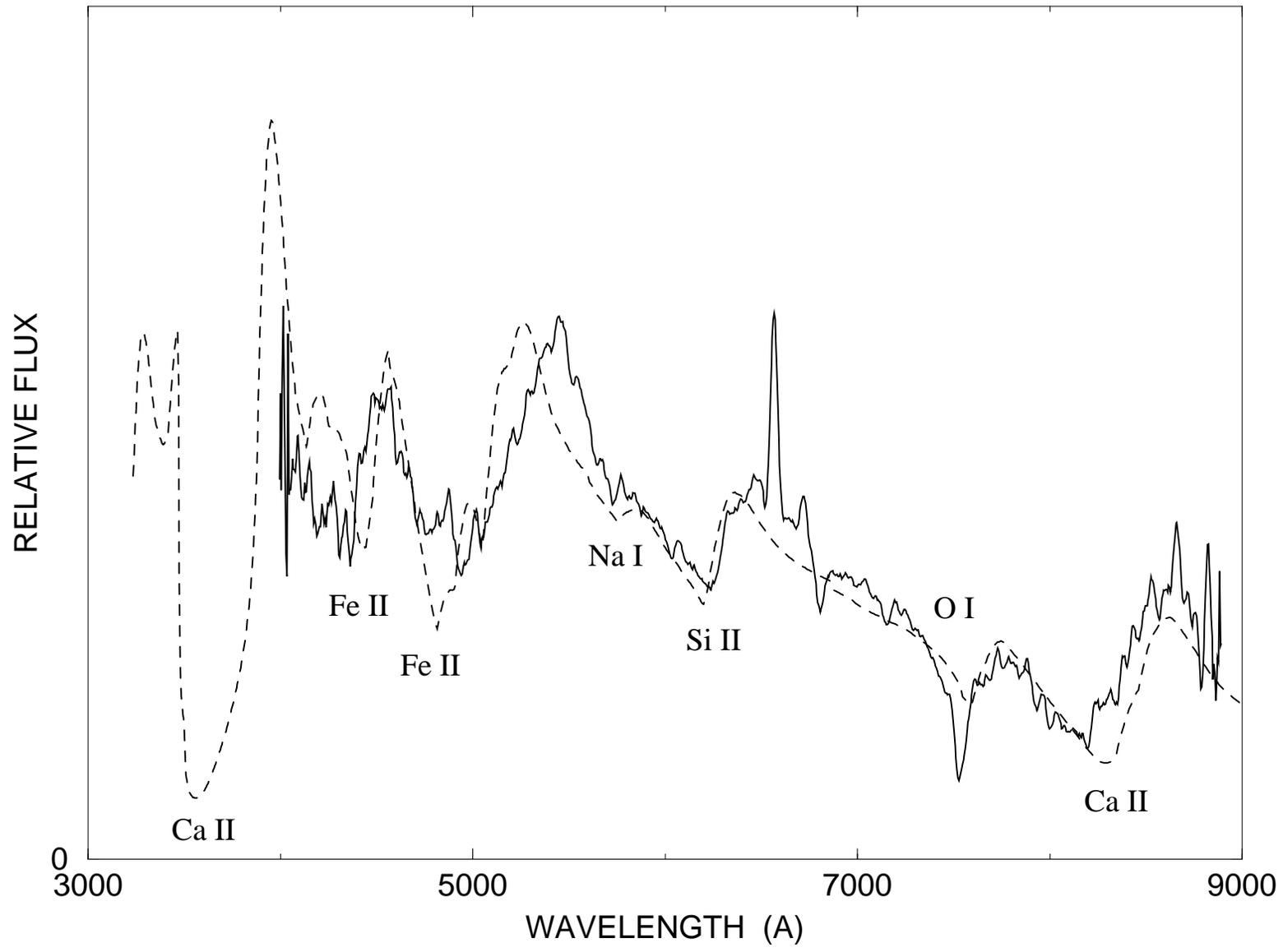
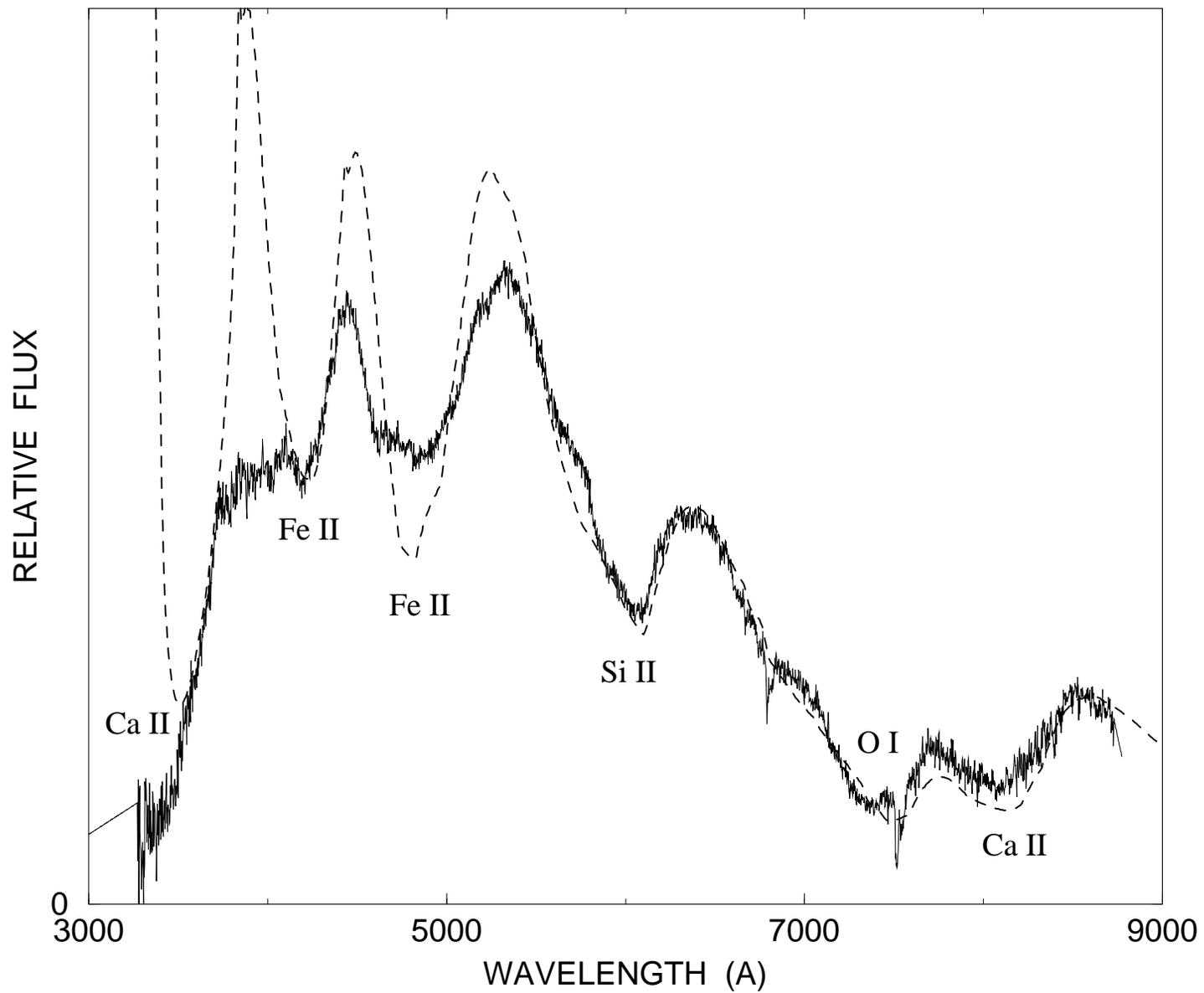
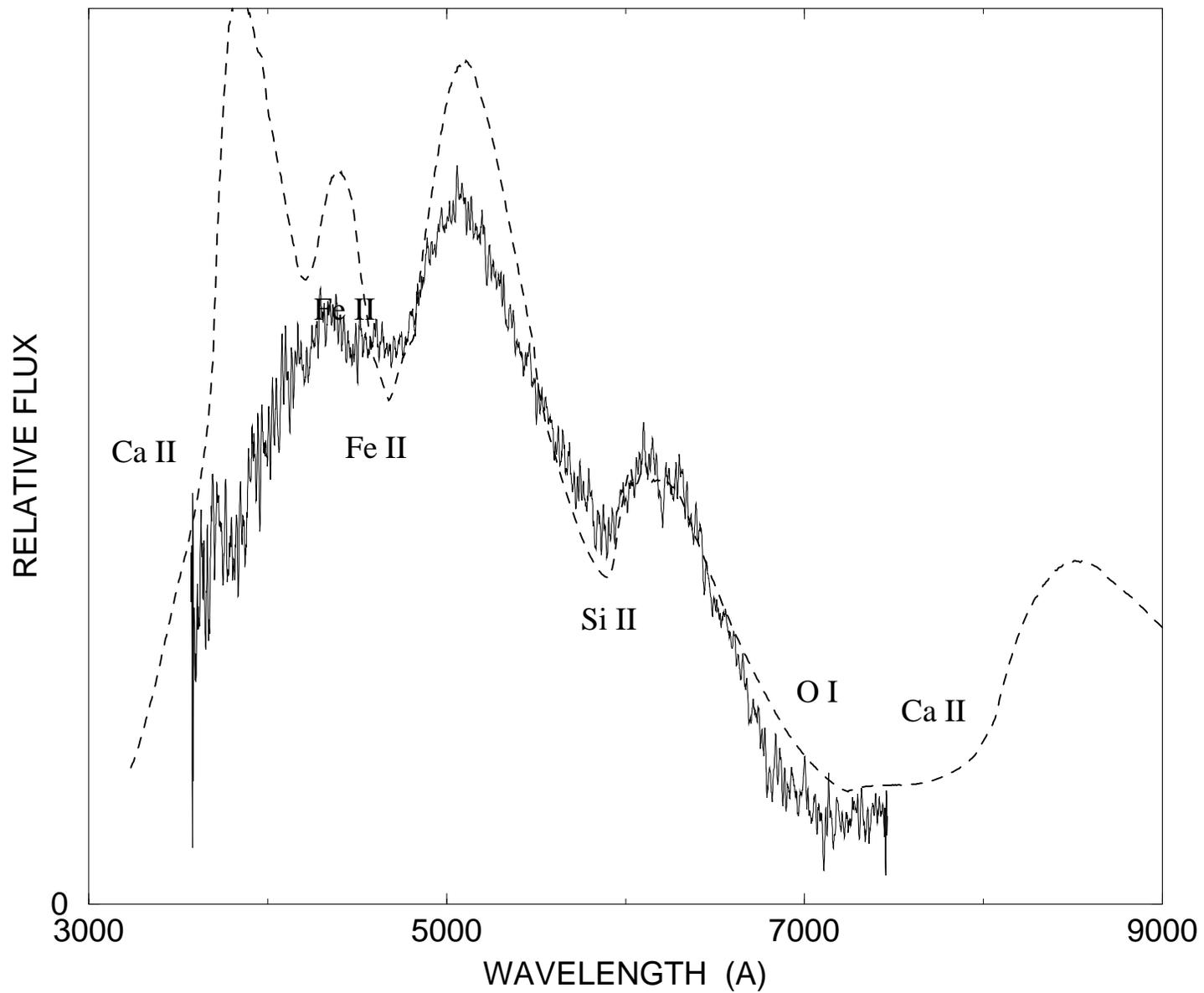
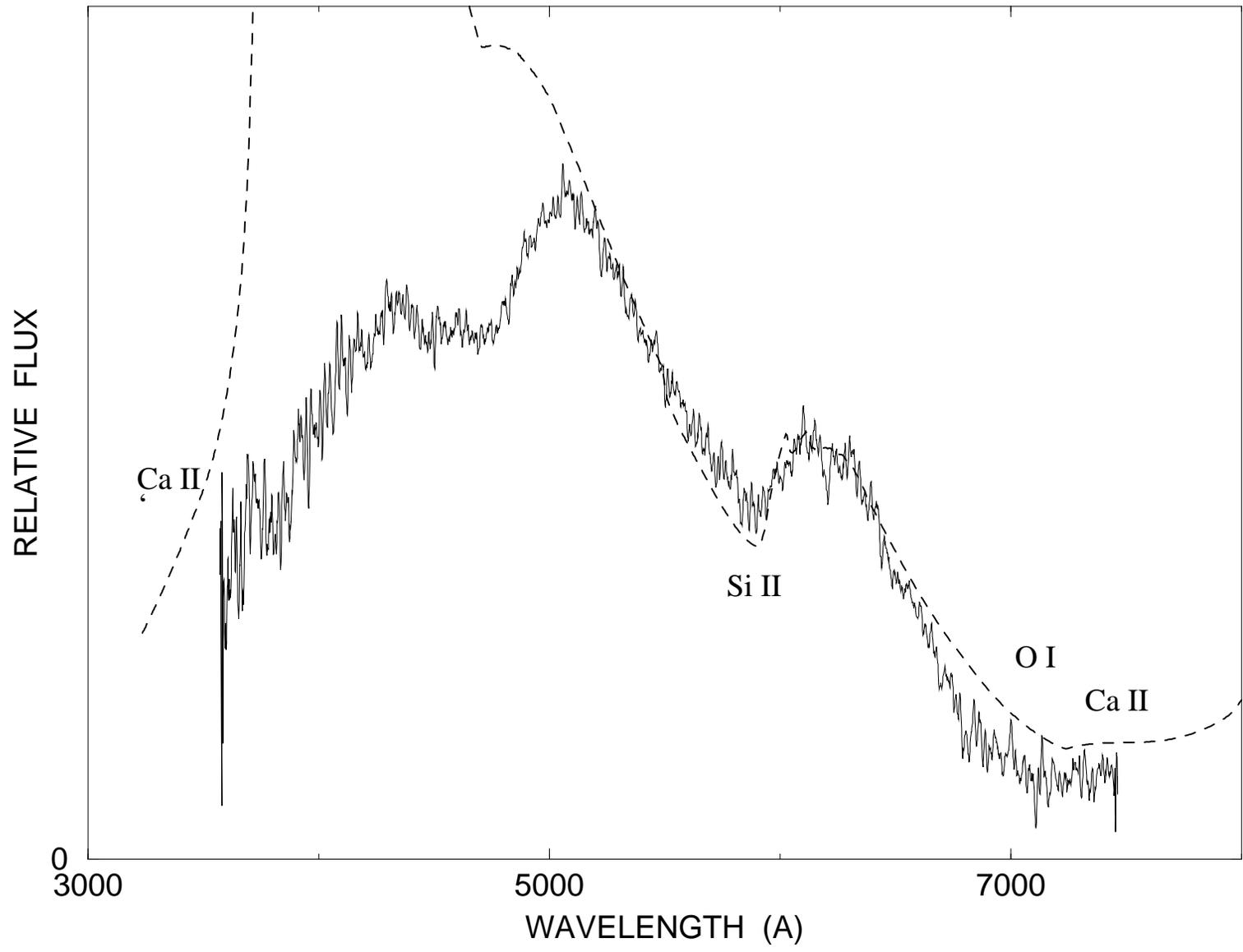
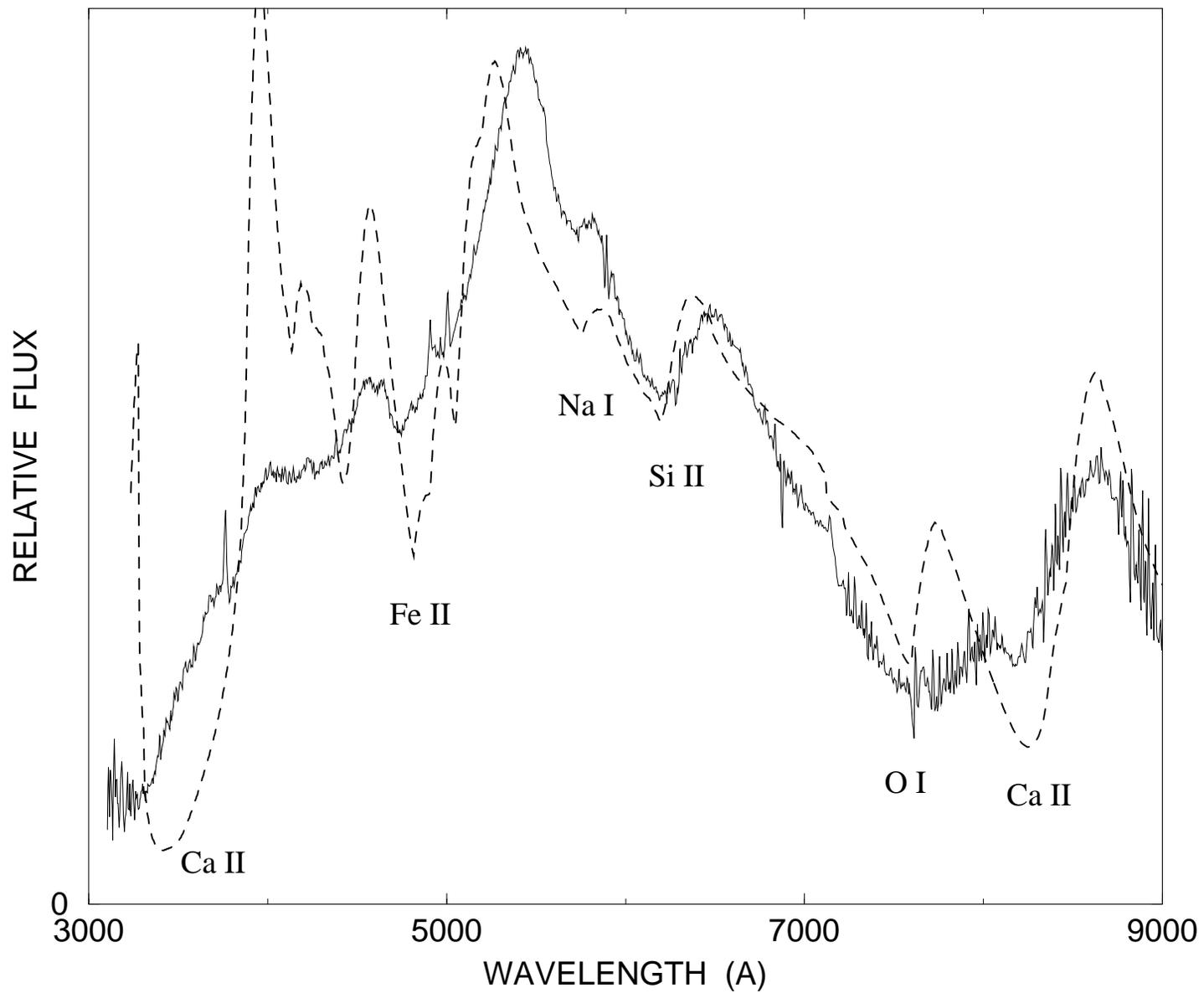
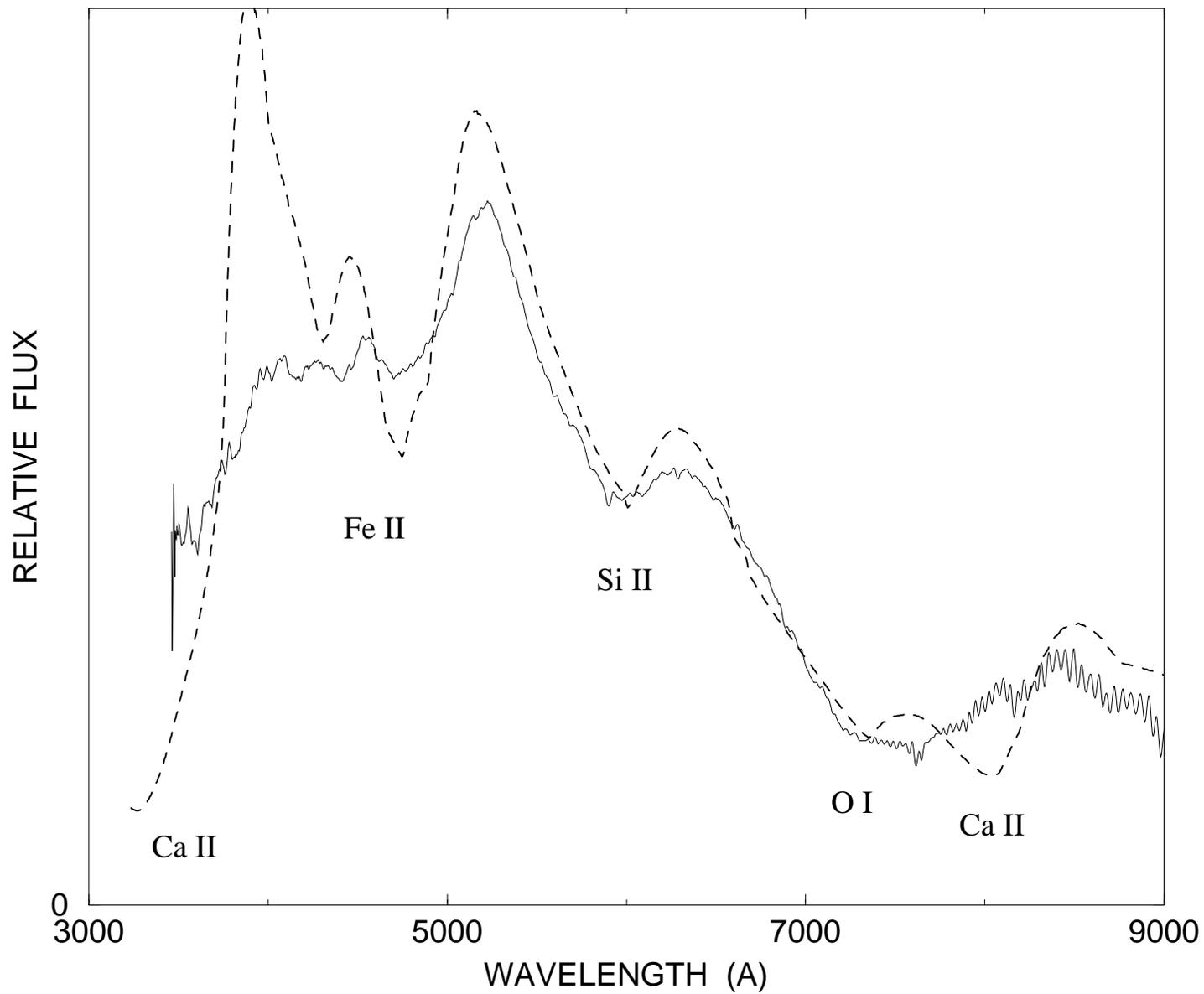
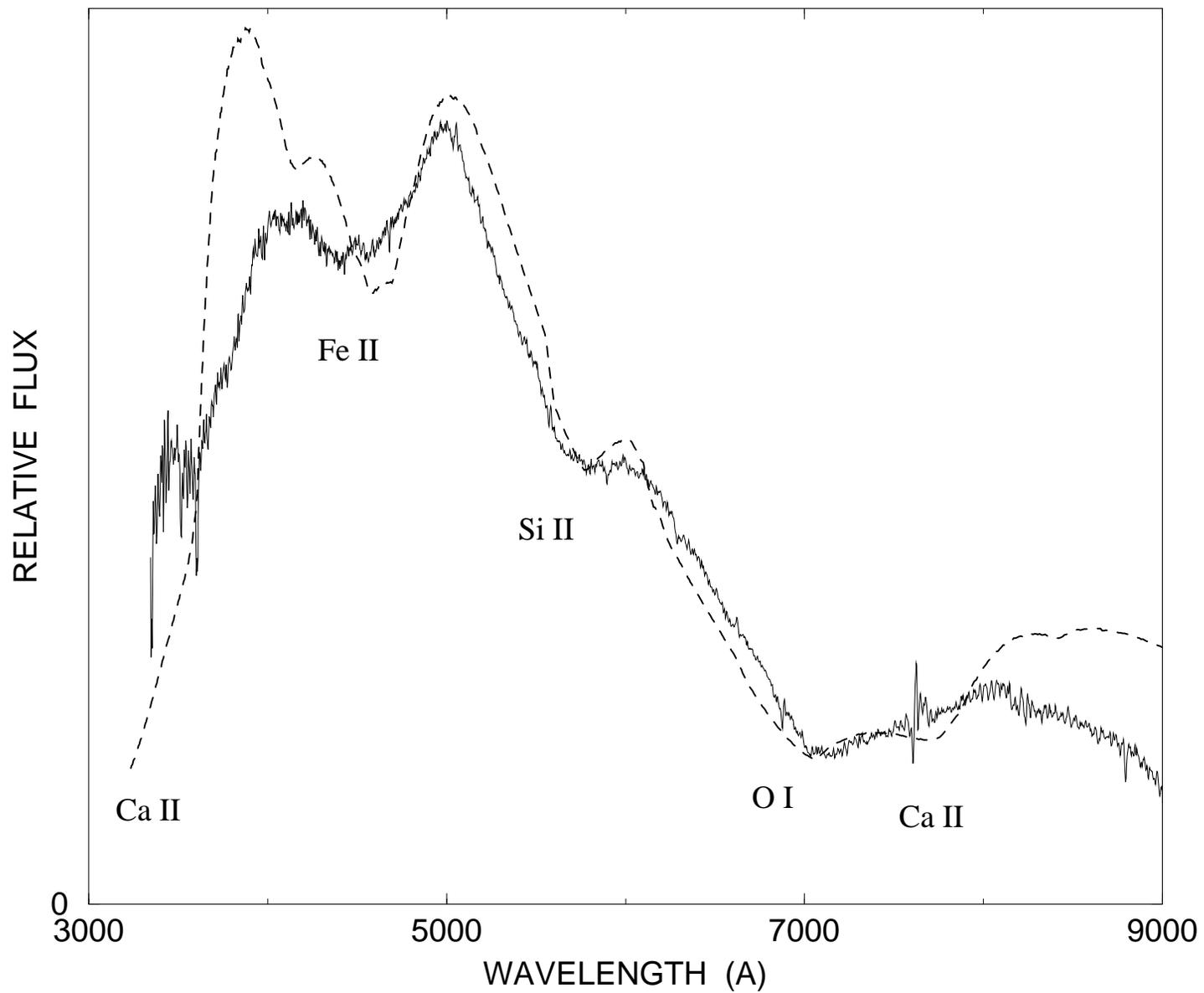
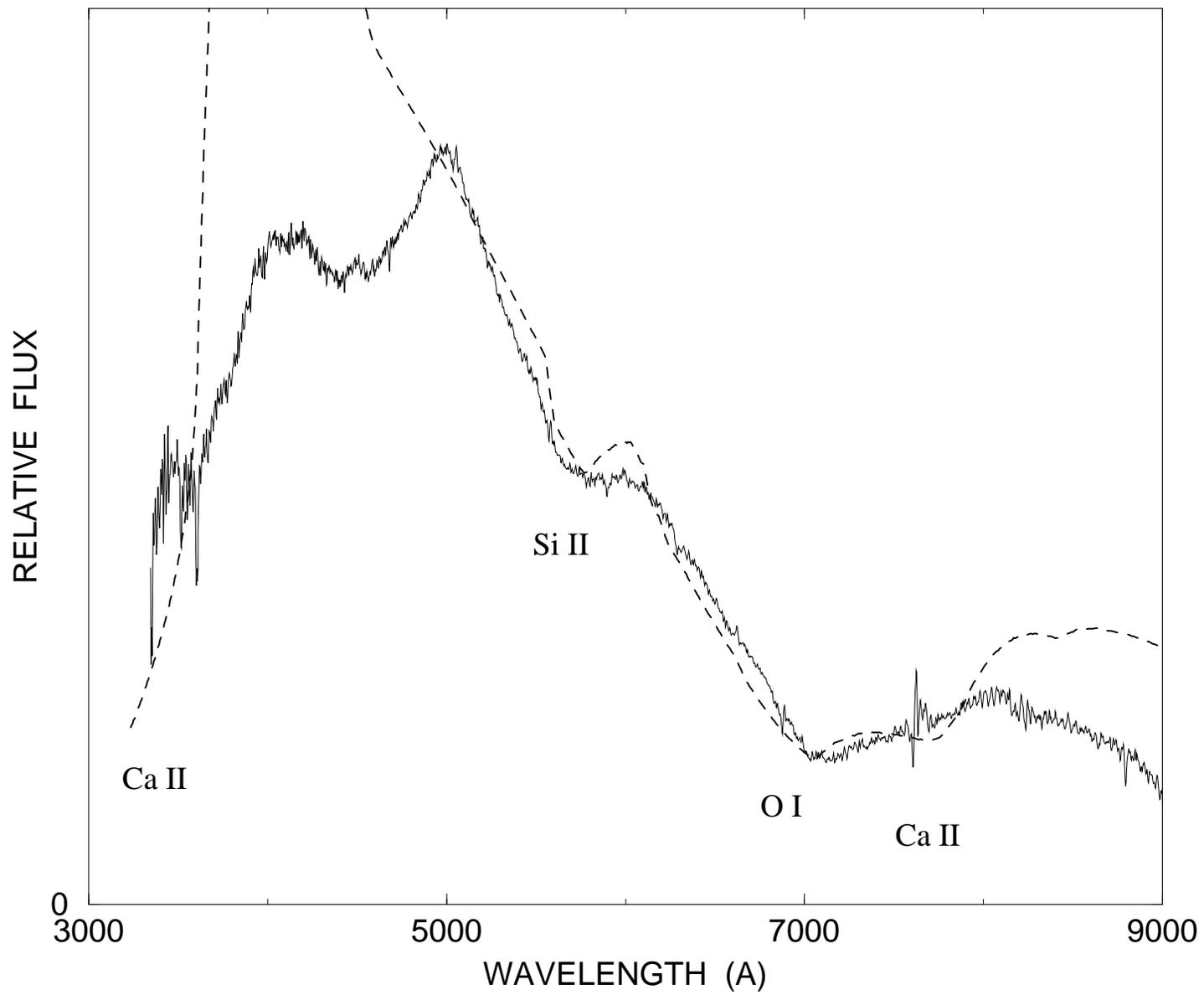
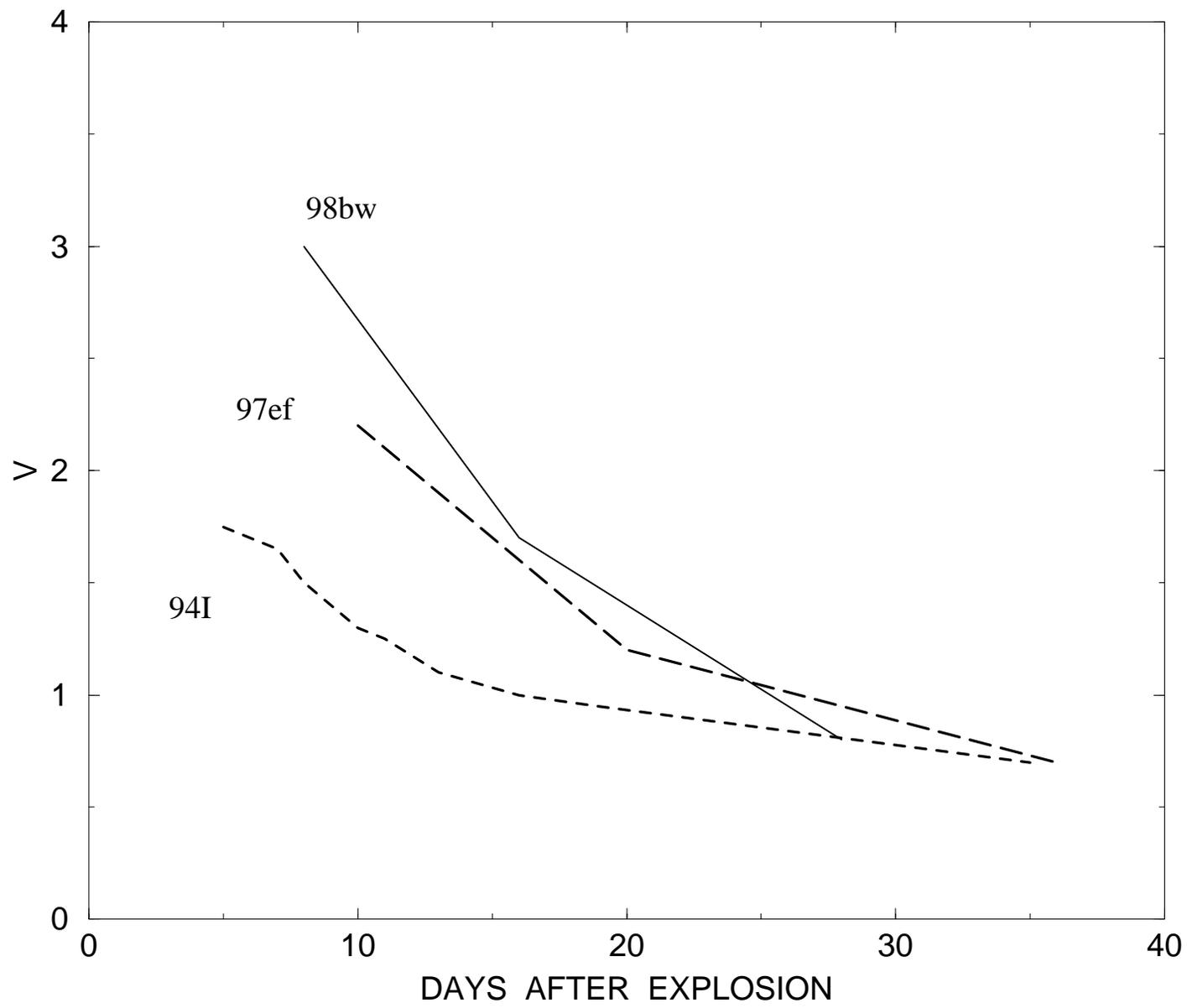
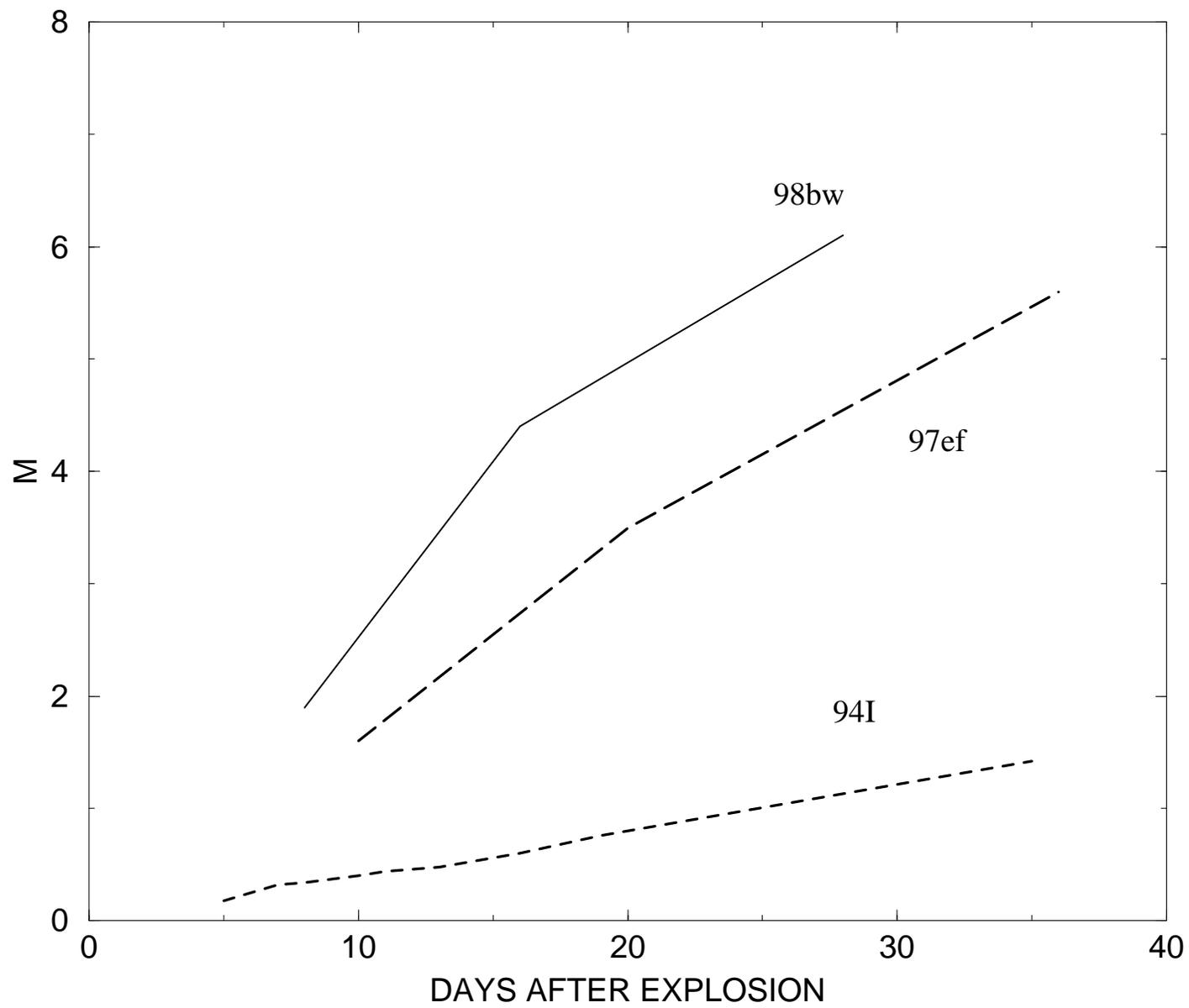
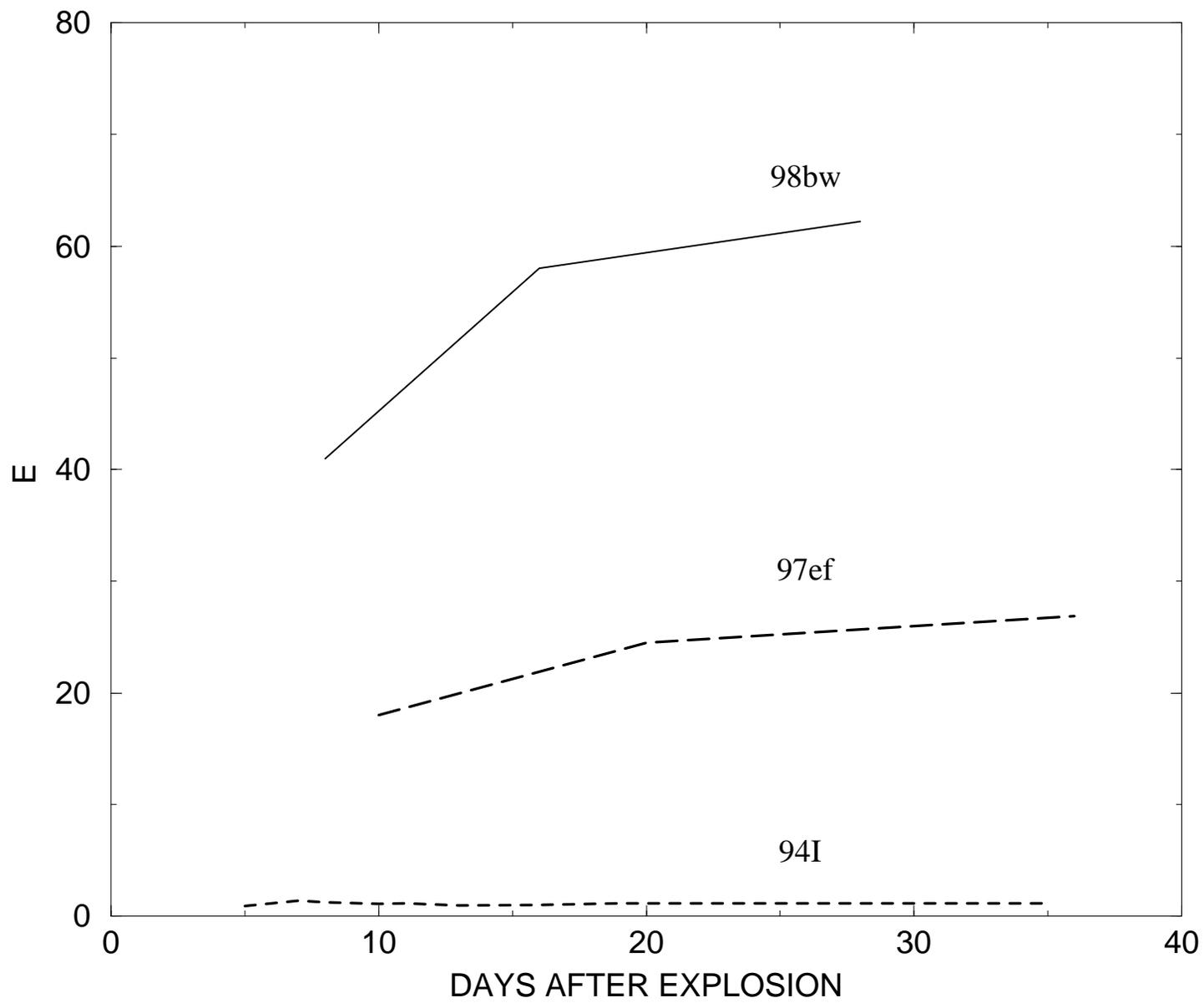
\end{document}